# Fabrication of nano-structured Titania - Thin-Films on Carbon-coated Nickel sheets

Wilfried Wunderlich, Nguyen Thi Hue, Sakae Tanemura,
Nagoya Institute of Technology, Dept. of Environmental Technology,
Gokiso Showa-ku, 466-8555 Nagoya, e-mail: wunder@system.nitech.ac.jp

**Abstract**
This research for Eco-materials focuses on the improvement of photo response of titania thin films. The goal of this research is to find a suitable substrate for decreasing the band-gap of titania, to improve the microstructure and to seek for new processing techniques in nano-meter scale. The sol-gel technique using titanium-tetra-iso-propoxide (TTIP), di-ethanol-amine (DEA) and ethanol was applied by using the dip-coating method. The substrate is a nickel-sheet covered with amorphous carbon-layers or carbon-nano-tubes (CNT), both produced with plasma-enhanced chemical vapor deposition (PECVD) by varying the temperature and the gas molecule velocity. The microstructure of these titania thin films show nanometer-structured pores, which are supposed to decrease the band gap according to the literature. Possible applications for photo catalytic devices will be discussed.

## 1. Introduction

Titanium dioxide is one of the most efficient materials as a photo catalyst or for photovoltaic devices, like dye-sensitized solar cells [1] or water purifying systems [2]. For further improvement of the optical response, e.g. to match the impinged solar maximum irradiance to the band-gap, we have examined the dependence of the band gap on the lattice constants [3-5]. Also it has been reported, that in the case of titania nano-particles the band gap becomes wider [6], while in the case of nano-porous materials it becomes narrower [7]. An analogous behavior, an increase of the surface energy compared to bulk value for nano-particles and an decrease for nano-pores was found in a study on $Al_2O_3$-MgO-ceramics [8]. This leads to the conclusion, that both structures, nano-particle and nano-pores, behave controversially. While nano-particles can be produced easily by many different processing methods, the search for suitable processing methods for nano-porous materials is still in progress. The inverse opal structure has been applied for producing nano-porous titania [9] and is the most promising method, because it has the advantage of exact control of the pore size, but rather costly. This paper suggests another fabrication method of nano-porous titania thin films by using carbon-nano-tubes (CNT) as nano-templates: The processing consists of two steps, growth of CNT as nano-templates and then subsequently sol-gel method is applied in order to achieve the nano-porous titania thin film.

After the discovery of the growth of CNT by plasma enhanced chemical vapor deposition (PECVD) using metallic catalyst like Pd, Ni, Fe as substrates [10], this technique was improved, so it became achievable to produce homogenously distributed CNT arrays on wide areas of a flat substrate. The metal surfaces act as a catalyst for the decomposition of the acetylen gas, leading to the growth of CNT [11], which are topped with a nano-size metallic particle. By changing the process parameters amorphous carbon layers can be produced with this technique and show also interesting photo-voltaic properties.

The sol-gel processing of titania has been already reported in the literature [12-14], also for producing thin films in a cost-efficient method [14]. The microstructure of these titania thin films consists of sintered nano-particles with lower density than advanced sputtering methods like rf helicon magnetron sputtering [4,15], but is sufficiently good for the desired application.

In this research we introduce a new approach to fabricate nano-porous titania thin films by the combination of nano-structured templates such as CNT or CL and the sol-gel-dip-coating method.

**Experimental**
Nickel sheets (99.99%, Nialco, Japan) were cleaned with ethanol and then deposited with carbon-nano-tubes (CNT) or carbon layers (CL) in a plasma-enhanced chemical vapor deposition (PECVD) chamber, where the flow of ammoniac and acetylene gas in a ratio of 2:1 decomposes into carbon and hydrogen with the aid of a plasma, the details are described elsewhere [10,11]. According to the processing parameters, temperature, electric field of the plasma, and the gas molecule velocity controlled by the inlet gas flow, the microstructure of the deposit changes from amorphous carbon-layers with wavy surfaces or thick or thin carbon-nano-tubes. The microstructure of the deposited layers was characterized by scanning electron microcopy (SEM) JEOL 5600 at 30kV and transmission electron microcopy (TEM) JEOL 3010 EX at 300kV. Whether carbon nano tubes or carbon layers are formed, is controlled by the processing parameter, especially the temperature controls the competition of CNT nucleation or CNT growth and only the optimum temperature of about $650^{o}$C leads to fine CNT with a minimum thickness of about 5nm and dense and almost homogeneous distribution.

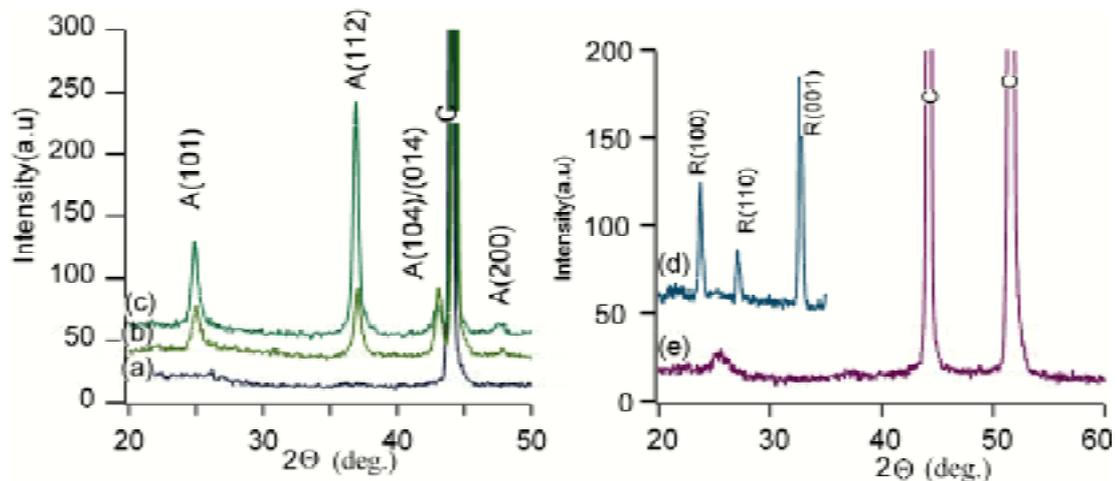

Fig. 1 X-Ray analysis of the titania layers on the a) CL substrate, b) anatase film one time dipped, c) anatase film three times dipped, d) rutile film, e) CNT substrate.

Using these Nickel-Carbon composites as a nano-sized template, thin films of titania were deposited by sol-gel technique. The precursor solution contains titanium-tetra-iso-propoxide (TTIP), ethanol (both from Wako Pure Chemical Ltd.), and di-ethanol-amine (DEA, Kishida Chemical Company) and 6h after mixing the hydrolysis reaction is completed. The specimens were dipped slowly in this liquid, removed, and dried vertically in air for 24h, allowing the gelation to occur. By different annealing temperatures the formation of either, anatase phase or rutile phase, can be controlled. This sequence of dip-coating and annealing was applied either one time or three times. The specimens were characterized by X-ray diffraction (XRD) Rigaku Rint2000 using Cu-K$\alpha$ and the JSPS data base for indicating the diffraction peaks.

**Results and Discussion**

The results of the XRD measurements are shown in Fig. 1. The peaks of the CL (Fig. 1 a) and the CNT (Fig. 1. e) on the Ni-sheets before dip-coating appear at $47^{o}$ and $55^{o}$, while a small peak at $27^{o}$ appears only clearly at the CNT-specimens. After dip-coating of the titania thin films, the carbon peaks are weaker, and additionally the titania peaks appear, which correspond very well to the data of the anatase and rutile phase. Specimens fired at moderate temperatures show at $25^{o}$ the anatase (101)- and at $37^{o}$ the anatase (112)-peak (Fig. 1 b, c), and specimens fired at higher temperatures show the following rutile peaks, (100) at $24^{o}$, (110) at $27^{o}$, (100) at $32^{o}$, respectively. Comparing specimens, dip coated one time (Fig. 1b) and three times (Fig. 1c), the peak heights increases, while the width decreases, indicating a more dense crystallization with less defects after subsequent coating. The microstructure as characterized by SEM is shown in Fig. 2 for specimens dip-coated only once. Patterns of thicker titania areas, in the following called wires, appear with a bright contrast for all four combinations, a) anatase on CL, b) rutile on CL, c) anatase on CNT and d) rutile on CNT. The areas in between the wires are either almost uncoated in the case of CL (Fig. 2a, b) or homogeneously coated with a smaller thickness (Fig. 2 d) or rather inhomogeneous coating (Fig. 2c) in the case of CNT substrates. The coating inhomogeneities in Fig. 2c (some parts show titania films instead of wires) can be explained by the roughness of the nickel sheets, like scratches etc. The connections between the wires at the triple points are rather sharp in the case of anatase (Fig. 2a, c), while they are more rounded for rutile (Fig. 2b, d), which can be explained by the faster diffusion at the higher annealing temperature. At the edges of the specimens, where the covering of the substrate by the dipping solution was less effective, the wires became thinner or consist of unconnected islands. The width and height of the titania wires in the completely covered areas are 1μm in the case of anatase (Fig. 3a) and 0.7μm in the case of rutile (Fig. 3b). The wires consist of sintered nano-particles. In the case of anatase the edges of the wire consist of an outstanding rim with an additional height of about 0.3μm and thickness of 0.1μm. The formation of the wires can be explained by the retraction of the liquid during the drying process: The surface energy keeps the liquid in the wire-shape, which is consolidated as a solid wire during the subsequent firing. Whether these sub-micrometer structured wires can be possibly used in applications, e.g. as new templates for further processing, remains an open question.

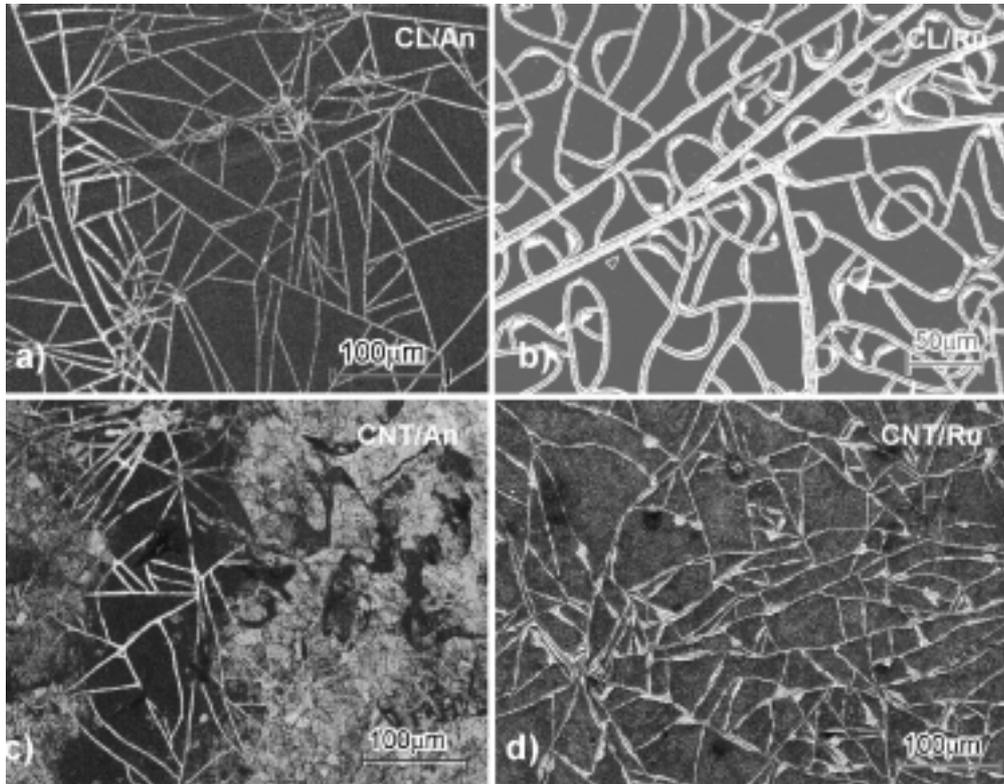

Fig. 2. Microstructure of a, c) anatase, b, d) rutile thin films on a, b) CL, c, d) CNT after one time dipping

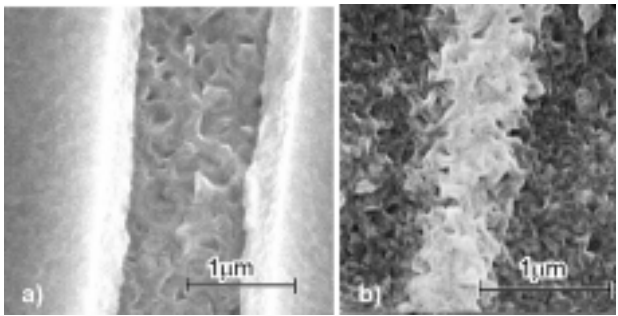

Fig. 3. Pattern forming wires, a) anatase , b) rutile

Specimens dip-coated three times show a homogenous microstructure of the titania thin films over large areas (Fig. 4). The surface is rough and consists of many pores, which shape depends on the substrate (CL or CNT) or the annealing condition. The roughness is large for anatase on CL (Fig. 4a) or rutile on CL after one time dip-coating (Fig. 4c), but rather round-shaped pores are observed for rutile on CL or CNT (Fig. 4b and e). The size as deduced from the SEM micrographs is in all specimens around 50nm, except in rutile on CNT around 100nm (Fig. 4d). The larger pore size and the rounder shape of the pore rims in the case of rutile can be explained by the enhanced diffusion at the higher annealing temperature instead of that for anatase. The results show, that nano-porous titania thin film composites can be manufactured and the pore size and rim curvature depend strongly on the microstructure of the CL or CNT template as further explained in the following.

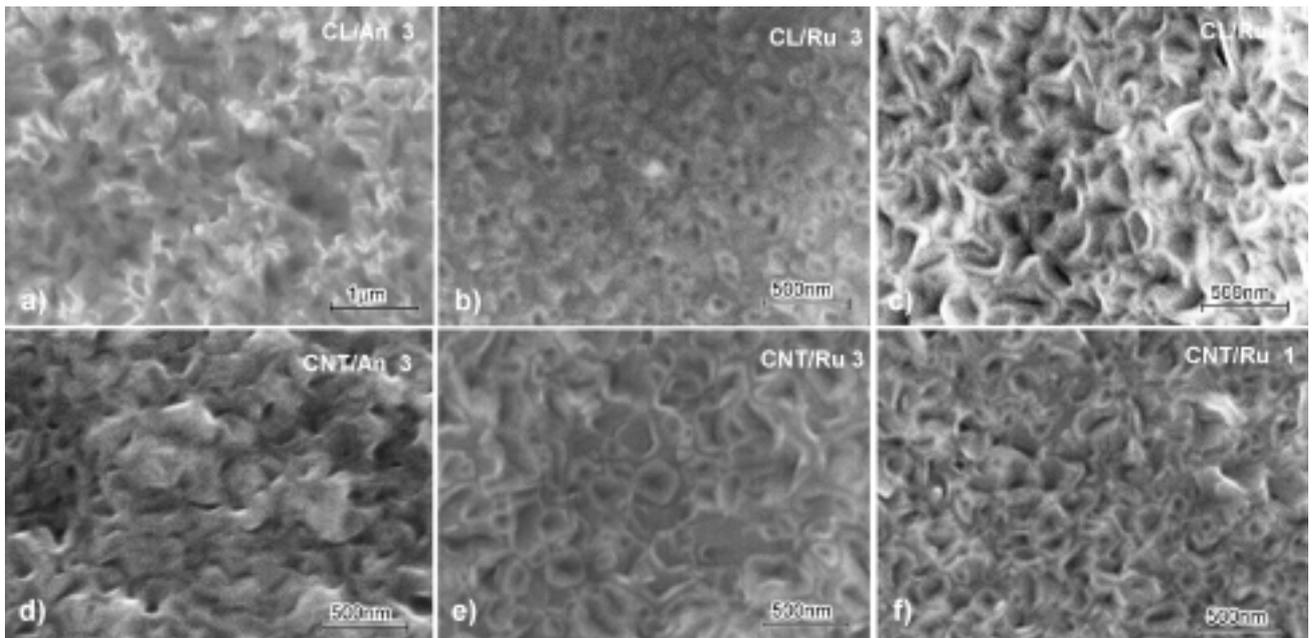

Fig. 4 Microstructure of the anatase (An) or rutile (Ru) thin film formed on different templates, a, b, c) CL, d, e, f) CNT; microstructure a,b, d,e) obtained after three time dipping, and c, f) after one time dipping showing the areas in between wires.

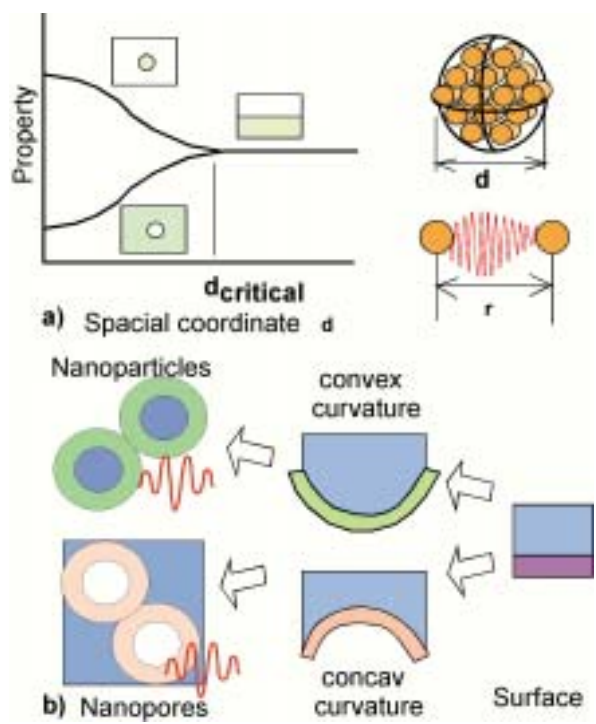

Fig. 5. Scheme showing a) the bifurcation of physical properties for nano-particles and nano-pores explained by b) the relationship between surfaces, nano-particles and nano-pores

The desirable band-gap of the titania films for the application in photo-active devices should be as narrow as possible. It is known from the literature that in the case of nano-porous titania in both phases, anatase and rutile, the band-gap is lowered [7], while for titania nano-particles the band-gap increases [6]. This seems to be a general physical principle, as illustrated in Fig. 5a: Concerning any physical property, like the band-gap in titania or the surface energy [8], there is a bifurcation, whether the structure consists of nano-particles or nano-pores. When the diameter $d$ of the particles or pores drops below a critical limit, the energy is increased for nano-particles or decreased for nano-pores. This controversial behavior can be explained by the fact, that the structural diameter $d$ of the pore or the particle becomes similar in size than the interaction distance between the atoms r (Fig. 5 a). The micro-structural feature, which distinguishes nano-pores from nano-particles, is the bending of the surface (Fig. 5 b). If a surface is bended in a convex shape, it becomes finally a particle; when it is bended concave, a pore is formed. The conclusion from this consideration is, that a buckling thin film, like the titania films in the present experiments, with high amount of strongly bended convex areas should behave similar as nano-particles, and those with high amount of concave bended areas should behave similar as nano-pores. The bending of the thin film presented in this paper, can be adjusted by the spacing of the CNT $s$, their thickness $d$, and the thickness of the titania thin film $t$, resulting in a pore radius $p = s\text{-}d\text{-}2t$ (Fig. 6). Strongly bonded areas of both types, concave and convex, can be obtained by narrow spacing $s$ and small $d$ (Fig. 6 a). Comparing the experimental micrographs, this is the case for anatase films on CNT, although the values for $s$ of about 50nm should be decreased further to less than 10nm, the limit, where the quantum size effects will occur. The case of

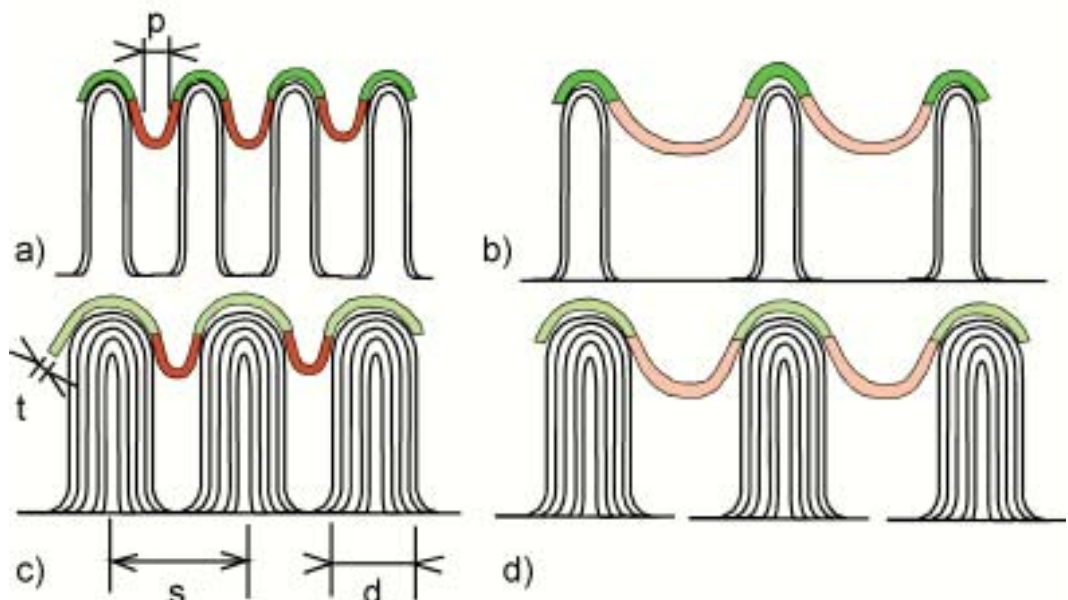

Fig. 6 Microstructural parameters influencing the bending and pore radius $p$ of the titania thin film deposited on the CNT nano-template, leading to the following curvature, a) strong convex, strong convave, b) strong convex, weak concave, c) weak convex, strong concave, d) weak convex, weak concave.

Fig. 6b has strongly bended convex areas, which can be easily realized in the experiment with a low density of thin CNT, similar like in the case rutile on CNT (Fig. 4 e). On the contrary, the case of Fig. 6 c has strongly bended concave areas, occurring in anatase on CL (Fig. 4 a), and finally the case Fig. 6 d, with weakly bended areas almost similar like conventional thin films. It is expected that only specimens with strongly bended concave areas behave like nano-porous titania, namely show a narrower bad gap and, hence, are promising materials for improving the efficiency of photovoltaic devices.

**Conclusions**

The new processing method described in this paper is a combination of nano-templates such as CNT or CL on nickel sheets prepared by PECVD and dip-coating of these templates into titania gels, which finally leads to a composite material, consisting of Ni-sheets covered with CL or CNT and nano-porous titania-thin films with a high amount of strongly bended pores. By varying the annealing temperature the formation of anatase or rutile thin films can be controlled. The areas of concave or convex bending can be adjusted by varying the thickness and spacing of the CNT on the nano-template. The obtained nano-porous titania films with narrowed band-gap will be candidate materials for photo-voltaic or photo-catalytic devices.

**Acknowledgement**

A part of this work was financially supported by the research program "21th century center of excellence (COE)", which is kindly acknowledged.